\documentclass[a4paper]{jpconf}
\usepackage{graphicx}
\usepackage{units}
\usepackage{amsmath}
\usepackage{amssymb}
\usepackage{footmisc}
\RequirePackage{mathptmx}

\renewcommand{\thefootnote}{\fnsymbol{footnote}}

\begin{document}
\title{First results on low-mass dark matter from the CRESST-III experiment}

\author{F Petricca$^1$, G Angloher$^1$, P Bauer$^1$, A Bento$^{1,8}$, C Bucci$^2$, L Canonica$^{2,9}$, X Defay$^3$, A Erb$^{3,10}$, F v Feilitzsch$^3$, N Ferreiro Iachellini$^1$, P Gorla$^2$, A G\"utlein$^{4,5}$, D Hauff$^1$, J Jochum$^6$, M Kiefer$^1$, H Kluck$^{4,5}$, H Kraus$^7$, J C Lanfranchi$^3$, A Lagenk\"amper$^3$, J Loebell$^6$, M Mancuso$^1$, E Mondragon$^3$, A M\"unster$^3$, C Pagliarone$^2$, W Potzel$^3$, F Pr\"obst$^1$, R Puig $^{4,5}$, F Reindl$^{4,5}$, J Rothe$^1$, K Sch\"affner$^{2,11}$, J Schieck$^{4,5}$, S Sch\"onert$^3$, W Seidel$^1$\footnote[2]{Deceased}, M Stahlberg$^{4,5}$, L Stodolsky$^1$, C Strandhagen$^6$, R Strauss$^1$, A Tanzke$^1$, H H Trinh Thi$^3$, C T\"urko\u{g}lu$^{4,5}$, A Ulrich$^3$, I Usherov$^6$, S Wawoczny$^3$, M Willers$^3$ and M W\"ustrich$^1$}

\address{$^1$ Max-Planck-Institut f\"ur Physik, F\"ohringer Ring 6, D-80805 M\"unchen, Germany}
\address{$^2$ INFN, Laboratori Nazionali del Gran Sasso, I-67010 Assergi, Italy}
\address{$^3$ Physik-Department E15, Technische Universit\"at M\"unchen, D-85747 Garching, Germany}
\address{$^4$ Institut f\"ur Hochenergiephysik der \"OAW, A-1050 Wien, Austria}
\address{$^5$ Atominstitut, Vienna University of Technology, A-1020 Wien, Austria}
\address{$^6$ Eberhard-Karls-Universit\"at T\"ubingen, D-72076 T\"ubingen, Germany}
\address{$^7$ Department of Physics, University of Oxford, Oxford OX1 3RH, United Kingdom}
\address{$^8$  Also at: LIBPhys, Departamento de Fisica, Universidade de Coimbra, P-3004 516 Coimbra, Portugal}
\address{$^9$  Also at: Massachusetts Institute of Technology, Cambridge, MA 02139, USA}
\address{$^{10}$ Also at: Walther-Mei\ss ner-Institut f\"ur Tieftemperaturforschung, D-85748 Garching, Germany}
\address{$^{11}$ Also at: GSSI-Gran Sasso Science Institute, 67100, L'Aquila, Italy}

\ead{petricca@mpp.mpg.de}

\begin{abstract}
The CRESST experiment, located at Laboratori Nazionali del Gran Sasso in Italy, searches for dark matter particles via their elastic scattering off nuclei in a target material. 
The CRESST target consists of scintillating CaWO$_4$ crystals, which are operated as cryogenic calorimeters at millikelvin temperatures. Each interaction in the CaWO$_4$ target crystal produces a phonon signal and a light signal that is measured by a second cryogenic calorimeter.
Since the CRESST-II result in 2015, the experiment is leading the field of direct dark matter search for dark matter masses below 1.7\,GeV/$c^2$, extending the reach of direct searches to the sub-GeV/$c^2$ mass region. 
For CRESST-III, whose Phase 1 started in July 2016, detectors have been optimized to reach the performance required to further probe the low-mass region with unprecedented sensitivity. 
In this contribution the achievements of the CRESST-III detectors will be discussed together with preliminary results and perspectives of Phase 1.
\end{abstract}

\section{Introduction}
The quest of the nature of dark matter is one of the most fundamental in today’s physics research. Although until today the existence of dark matter is inferred by gravitational effects only, one of the favoured solutions is the existence of new massive particles with an interaction cross section of the weak scale. 
Direct dark matter searches exploit a great variety of different detector technologies, mainly aiming to observe dark matter particles via their elastic scattering off nuclei in their detectors. The small energy deposit by the interaction in the detector material and the very small expected event rates represent the major challenges of these experiments.\\
Cryogenic experiments currently provide the best sensitivity for light dark matter particles, with the CRESST experiment advancing to the sub-GeV/$c^2$ dark matter particle mass regime.\\
The CRESST target consists of scintillating CaWO$_4$ crystals operated as cryogenic calorimeters at millikelvin temperatures ({\it phonon detectors}). Most of the energy deposited in a crystal by a particle interaction induces a heat signal, yielding a precise energy measurement. A small fraction of the deposited energy is emitted as scintillation light that is measured by a secondary independent cryogenic calorimeter ({\it light detector}), allowing for particle identification \cite{Angloher:2004tr, Angloher:2008jj}. A phonon detector and the corresponding light detector form a {\it detector module}. 
Both, phonon and light detectors are read out via tungsten transition edge sensors (TES). Each detector is equipped with a heater to stabilize the temperature in the operating point and to inject pulses which are needed for the energy calibration.\\
The experiment is based in the LNGS (Laboratori Nazionali del Gran Sasso) underground laboratory in central Italy.

\section{CRESST-III Detectors}
\begin{figure}[h]
	\begin{center}
	\begin{minipage}{15pc}
		\includegraphics[width=\linewidth]{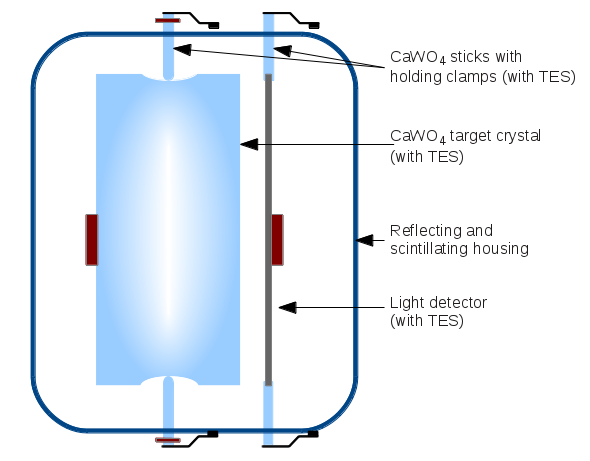}
	\end{minipage}\hspace{2pc}
	\begin{minipage}{14pc}
		\includegraphics[width=\linewidth]{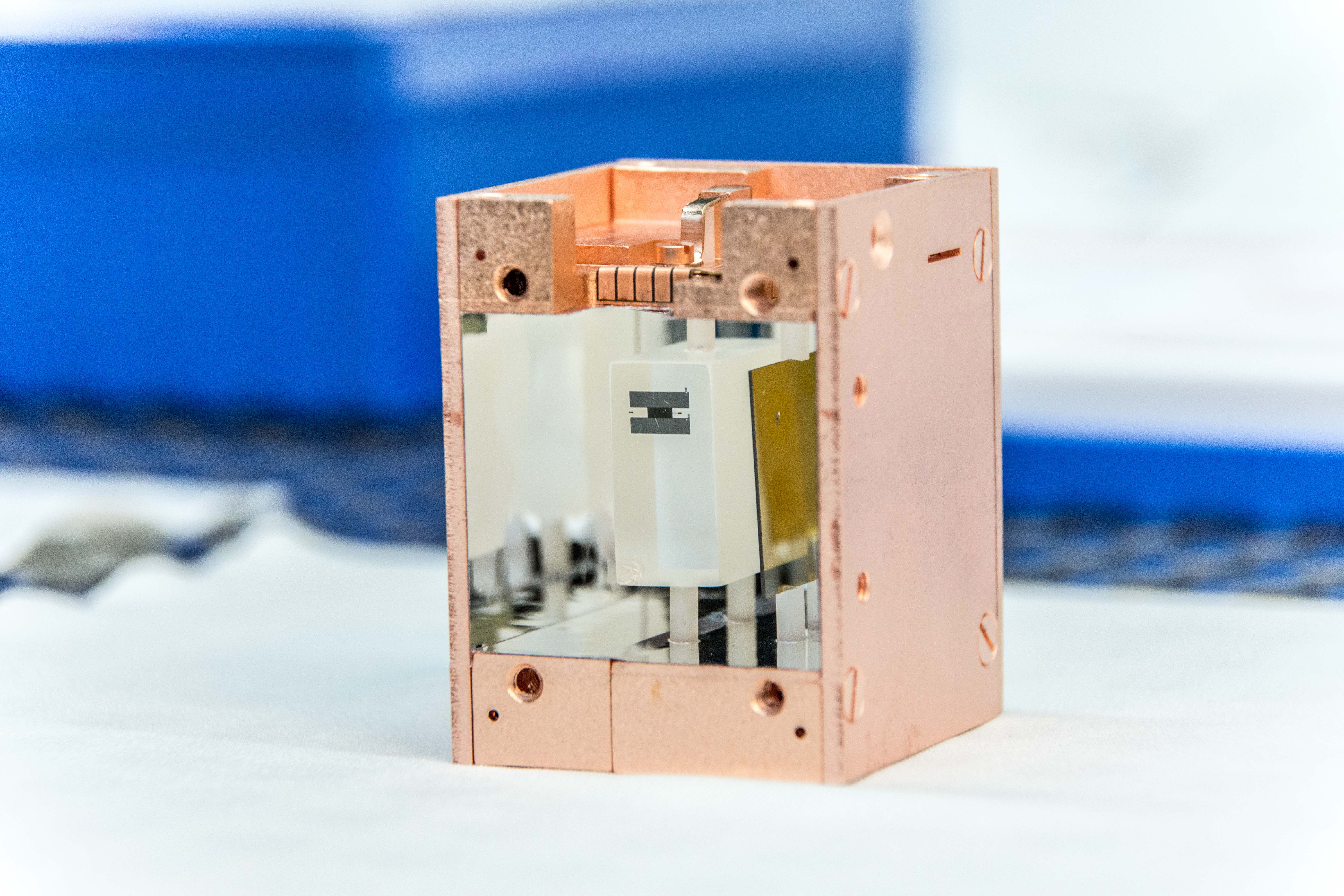}
	\end{minipage}
	\caption{\label{fig:small_module}Left: Schematic view of the detector design for CRESST-III modules. Right: Open CRESST-III detector module during mounting in the Gran Sasso facility.}
	\end{center}
\end{figure}

The results from CRESST-II Phase 2 \cite{Angloher:2015ewa, Angloher:2014myn} clearly demonstrate that the energy threshold is the main driver for low-mass dark matter search. Given the  kinematics of coherent, elastic dark matter particle-nucleus scattering, extremely low thresholds of $\mathcal{O}$(50\,eV) are necessary to access dark matter particle masses of $\mathcal{O}$(0.1\,GeV). Therefore, the detector optimization for CRESST-III followed a straight-forward approach thoroughly discussed in \cite{Angloher:2015eza}. The CaWO$_4$ target crystals of already available quality (i.e. radiopurity and optical properties) have been scaled down in mass from $\sim300$\,g to 24\,g and the thermometer design has been optimised to achieve a threshold of $<$100\,eV.\\
A schematic view of the detector layout is presented in fig.~\ref{fig:small_module}.

As can be seen from the picture in fig.\,\ref{fig:small_module}, the phonon and light detectors are held by CaWO$_4$ sticks, a design which was already successfully implemented in CRESST-II Phase 2 \cite{Angloher:2014myn, Strauss:2014hog}. In addition, a novel concept for an instrumented detector holder is introduced in which the sticks holding the phonon detectors are equipped with TESs which realize a veto against events related to the crystal support.

\section{CRESST-III Phase 1}
The first data-taking period of the CRESST-III experiment, referred to as Phase 1, has started in July 2016 and since November 2016 the experiment is steadily collecting dark matter data.\\
In the following we will briefly report on the first results obtained with the detector module Det-A which shows the best overall performance among the CRESST-III modules. For this first analysis a conservative analysis threshold of 100\,eV, corresponding to the original threshold goal, was chosen. 

\subsection{Data set and data analysis}
\begin{figure}[h]
	\begin{center}
	\begin{minipage}{17pc}
		\includegraphics[width=\linewidth]{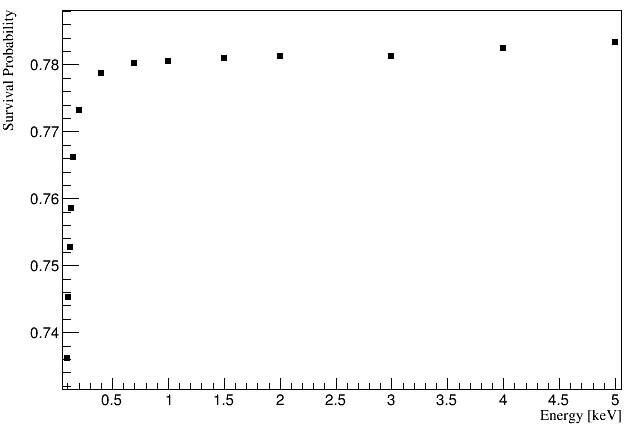}
		\caption{\label{fig:survival_probability}Signal survival probability between the analysis threshold of 100\,eV and 5\,keV for Det-A after cumulative application of all the selection criteria. The simulated pulses (data points) correspond to nuclear recoil events at discrete energies.}
	\end{minipage}\hspace{2pc}
	\begin{minipage}{18.5pc}
		\includegraphics[width=\linewidth]{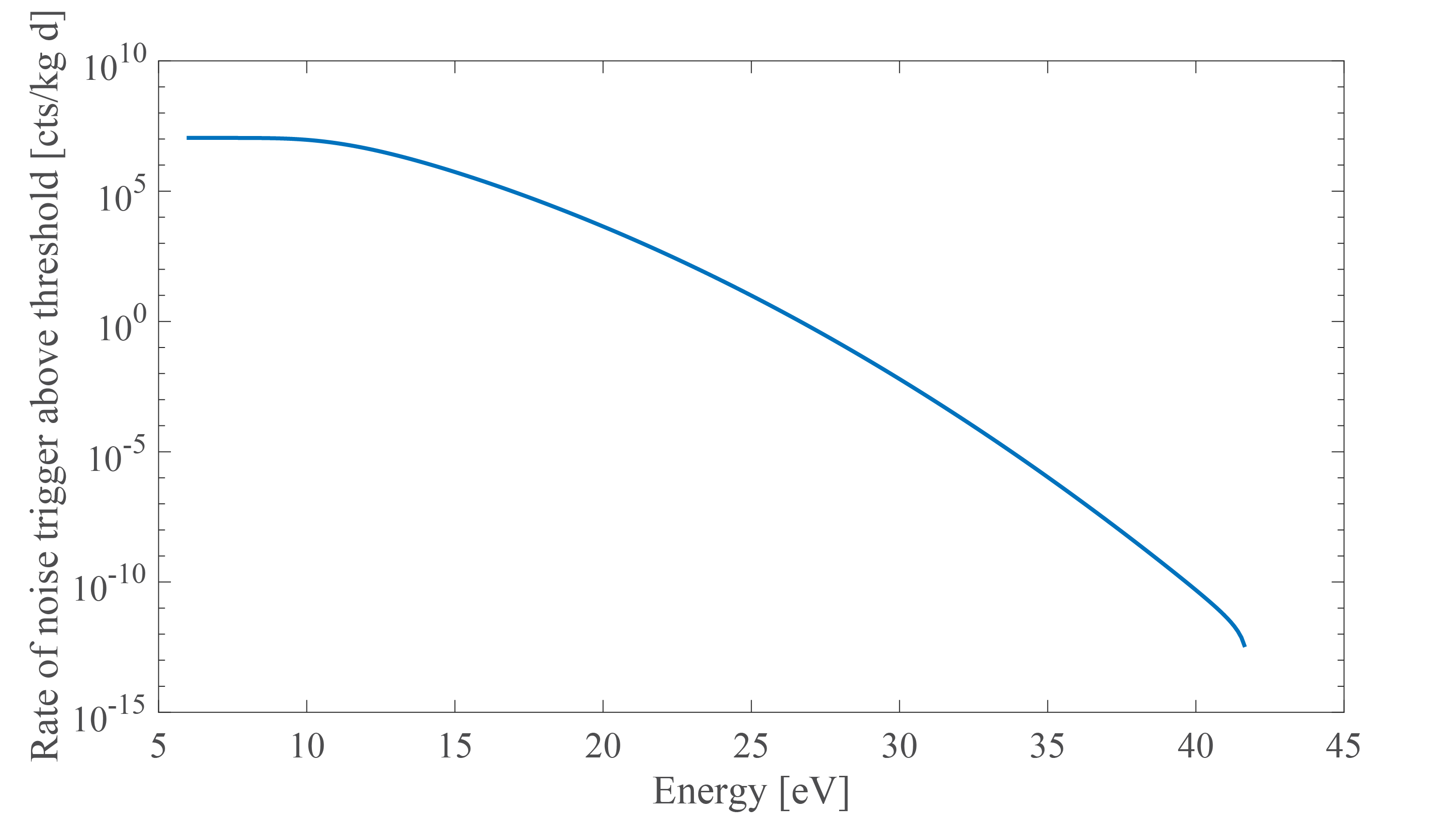}
		\caption{\label{fig:DetA_NTR}Noise trigger rate above threshold as function of threshold calculated for Det-A. For this detector, threshold values of 26.1 and 22.6\,eV correspond respectively to 1 and 100 noise triggers per (kg day) above threshold.}
	\end{minipage} 
	\end{center}
\end{figure}

The methods used for the analysis of the data are analogous to the ones thoroughly described and discussed in \cite{Angloher:2014myn, Angloher:2015ewa} and references therein. A blind analysis is carried out by first defining a statistically insignificant part of the data set as a training set. The few quality cuts that need to be applied to the raw data to ensure that only valid events are considered for further analysis are adjusted on the training set and then applied blindly (i.e. without any change) to the final data set. The training set is excluded from the final exposure.\\
For all cuts energy dependent efficiencies are measured by applying the cuts on a set of artificial nuclear recoil events with discrete known energies, created by superimposing signal templates on empty baselines periodically sampled throughout the run.\\
The fraction of signals with a certain simulated energy passing the cuts yields the respective survival probability. In fig.~\ref{fig:survival_probability} we show the cumulative signal survival probability for Det-A after all selection criteria are applied. The detector has a signal survival probability of 73.6\% at the analysis threshold of 100\,eV.

\subsection{Threshold determination}
To fully exploit the potentialities of the CRESST-III modules, a data acquisition system which continuously records the detectors has been implemented, allowing the off-line application of an optimised software trigger.\\ 
For CRESST-III we developed a method based on the optimum filtering technique to quantify the lowest trigger threshold achievable as a function of the acceptable amount of noise events. This approach sets a good base to define the value of the energy threshold considering the acceptable rate of misidentified noise events. Applying this method, which is thoroughly described in \cite{Mancuso}, we obtain for Det-A the noise trigger rate above threshold as a function of threshold shown in fig.\,\ref{fig:DetA_NTR}.

\subsection{Results and discussion}
\begin{figure}[h!]
	\begin{center}
		\begin{minipage}{17pc}
			\includegraphics[width=\linewidth]{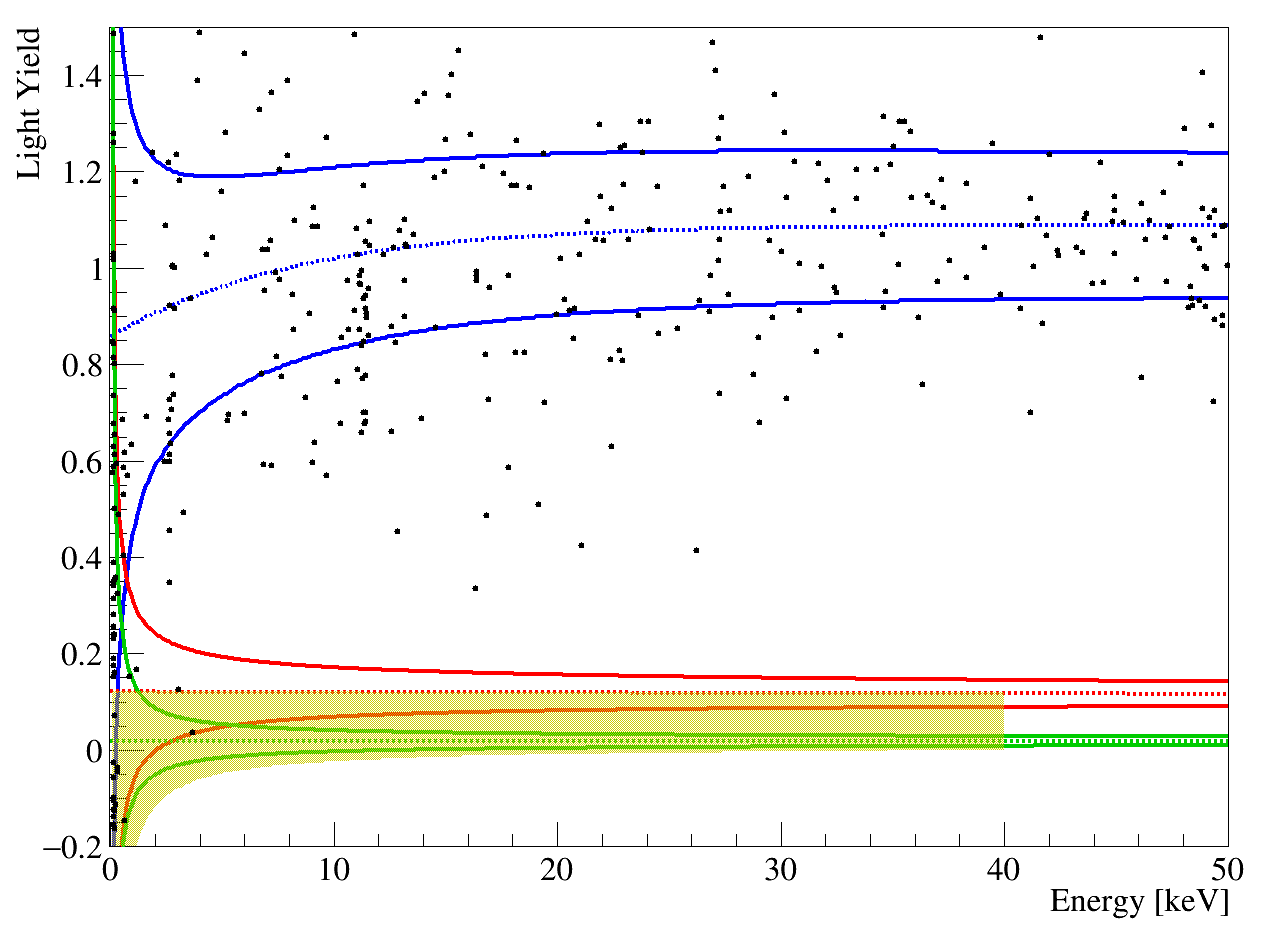}
			\caption{Data taken with Det-A in the light yield - energy plane. The solid lines mark the 90\% upper and lower boundaries of the e$^−$/$\gamma$-band (blue), the band for recoils off oxygen (red) and off tungsten (green). The dark matter acceptance region is highlighted in yellow.}\label{fig:DetA_LYwith_bands}
		\end{minipage}\hspace{2pc}
		\begin{minipage}{18.5pc}
			\includegraphics[width=\linewidth]{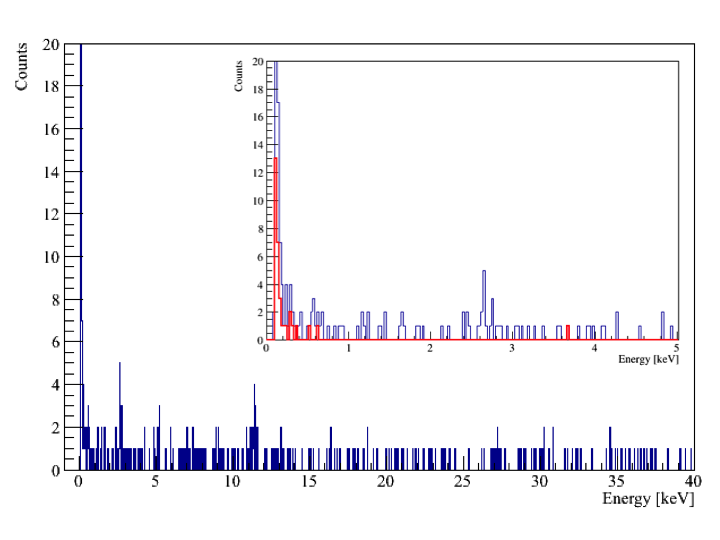}
			\caption{Low-energy spectrum of all events recorded with Det-A. The insert shows a zoom into the energy spectrum of all events (blue). Shown as red histogram are the events in the acceptance region (shaded yellow area in fig.\,\ref{fig:DetA_LYwith_bands})}\label{fig:DetA_spectra}
		\end{minipage} 
	\end{center}
\end{figure}

All events surviving the selection criteria, corresponding to \unit[2.39]{kg days} of raw data taken with Det-A, are presented in fig.~\ref{fig:DetA_LYwith_bands} in the light yield-energy plane. The light yield is defined for every event as the ratio of light to phonon signal. Electron recoils have a light yield set to one by calibration (at the calibration energy of $\sim$\unit[63]{keV}). Nuclear recoils produce less light than electron recoils, the reduction being quantified by the quenching factors for the respective target nuclei, which are precisely known from dedicated independent measurements \cite{Strauss:2014zia}. In fig.~\ref{fig:DetA_LYwith_bands} the solid blue lines mark the \unit[90]{\%} upper and lower boundaries of the e$^-/\gamma$-band, with \unit[80]{\%} of electron recoil events expected in between. From this band, with the knowledge of the quenching factors for the different nuclei present in the target material, the nuclear recoil bands for scatterings off tungsten, calcium and oxygen (respectively solid green, not shown and solid red in fig.~\ref{fig:DetA_LYwith_bands}) are calculated analytically.
The acceptance region (yellow region in fig.\,\ref{fig:DetA_LYwith_bands}) is defined to extend in energy between the analysis threshold of 100\,eV and 40\,keV and in light yield to span from the 99.5\% lower boundary of the tungsten band to the center of the oxygen band, in agreement with previous analyses.

\begin{figure}[h!]
	\begin{center}
		\includegraphics[width=.9\linewidth]{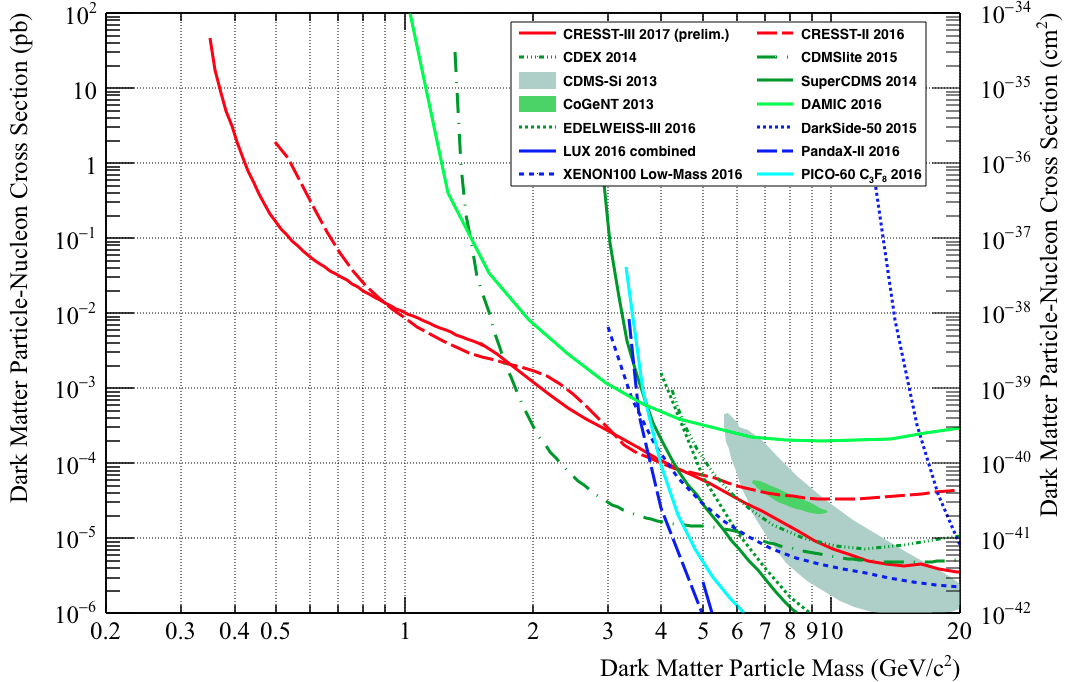}
	\end{center}
	\vspace*{-3mm}
	\caption{Parameter space for elastic spin-independent dark matter-nucleon scattering. The first result from CRESST-III Phase 1 (solid red) is compared with the limit from CRESST-II Phase 2 (dashed red) \cite{Angloher:2015ewa}. For comparison, exclusion limits (90\% C.L.) of other dark matter experiments are shown \cite{Agnese:2015nto, Agnese:2014aze, Aguilar-Arevalo:2016ndq, Agnes:2014bvk, Tan:2016zwf, Amole:2017dex, Yue:2014qdu, Hehn:2016nll, Akerib:2016vxi, Aprile:2016wwo}. The favoured parameter space reported by CDMS-Si \cite{Agnese:2013rvf} and CoGeNT \cite{Aalseth:2012if} are drawn as shaded regions.}
	\label{fig:LimitPlot}
\end{figure}

A total of 33 events are observed in the acceptance region, clustering at energies from threshold to about 500\,eV, with one outlier at 3.7\,keV. The energy spectra of the events in the acceptance region (red) and of all events observed in Det-A (blue) are presented in fig.\,\ref{fig:DetA_spectra}.
We use the Yellin optimum interval method \cite{Yellin2002_Limit, optimum_I} to calculate an upper limit with \unit[90]{\%} confidence level on the elastic spin-independent interaction cross-section of dark matter particles with nucleons. The exclusion limit resulting from the blind analysis reported here is drawn in solid red in fig.\,\ref{fig:LimitPlot}.

Compared to the previous CRESST result \cite{Angloher:2015ewa} shown in fig.\,\ref{fig:LimitPlot} as dashed red line, the reach is extended down to dark matter particle masses of 350\,MeV/c$^2$. At 500\,MeV/c$^2$ the sensitivity is improved by one order of magnitude, while for higher masses the limit is comparable to that of the previous result, despite the very small exposure.\\
The steep rise of the particle event rate towards lower energies observed in the acceptance region below $\sim$500\,eV, the origin of which is presently unknown, hinders the sensitivity for low-mass searches (i.e. dark matter particle masses below $\sim$1\,GeV/c$^2$). The presence of the event at 3.7\,keV is the cause of the limited sensitivity in the intermediate mass range (i.e. between $\sim$1\,GeV/c$^2$ and $\sim$5\,GeV/c$^2$). At higher masses the sensitivity scales, as expected, with exposure.

\section{Conclusion and outlook}
The first results on low-mass dark matter obtained with the Phase 1 of CRESST-III confirm that a low energy threshold represents a crucial requirement for direct dark matter searches aiming to achieve sensitivity to dark matter particles with masses in the \unit[1]{GeV/$c^2$} range and below.\\
With only \unit[2.39]{kg days} of raw data taken with Det-A and with an analysis threshold conservatively set at 100\,eV, the CRESST-III experiment improves the sensitivity of CRESST-II by one order of magnitude for a dark matter particle mass of 500\,MeV/c$^2$ and further extends the reach of the experiment down to 350\,MeV/c$^2$, reaffirming its leading sensitivity for light dark matter.\\
Despite the presence of background in the acceptance region, using the full exposure of the CRESST-III experiment and further extending the lower boundary of the search energy range down to the detector threshold, a significant progress is expected in the near future in the exploration of the low-mass regime.


\section*{References}
\bibliographystyle{iopart-num}
\bibliography{Petricca_DarkMatter}

\end{document}